\documentclass{mecbic}
\providecommand{\anno}{2011} \providecommand{\firstpage}{5}
\providecommand{\lastpage}{6}
\usepackage{breakurl}

\usepackage{color}
\usepackage{amsmath}
\usepackage{amssymb}

\title{Petri Nets and Bio-Modelling\\ {\large and how to benefit from their synergy}}

\author{Jetty Kleijn \institute{LIACS, Leiden University, 2300 RA, The Netherlands} \email{kleijn@liacs.nl} \and
Maciej Koutny\institute{School of Computing Science,Newcastle University,\\
Newcastle upon Tyne, NE1 7RU, United Kingdom} \email{maciej.koutny@ncl.ac.uk}
\and Grzegorz Rozenberg \institute{Leiden Center for Natural Computing, Leiden University\\
Niels Bohrweg 1, 2333 CA Leiden, The Netherlands} \email{rozenber@liacs.nl}}

\def\titlerunning{Petri Nets and Bio-Modelling}
\def\authorrunning{J. Kleijn, M. Koutny and G. Rozenberg}

\begin{document}

\maketitle

\pagestyle{plain}
\pagenumbering{arabic}
\setcounter{page}{5}

\begin{abstract}

\end{abstract}

Petri nets are a general, well-established model of concurrent and distributed
computation featuring a wealth of tools for the analysis and verification of
their behavioural properties. 
On the other hand, to understand specific biological processes 
different formalisations have been proposed. 
Examples here are membrane systems and reaction systems
which are close abstractions of the functioning of the living cell.
Membrane systems are a computational model inspired by the way chemical
reactions take place in cells that are divided by membranes into
compartments. The central idea 
behind reaction systems is that the functioning of a living cell
is based on interactions between (a large number of) individual reactions, 
and moreover
these interactions are regulated by two main mechanisms:
facilitation and inhibition.
 
In this talk we are concerned with the intrinsic 
similarities and differences between Petri nets on the one hand, and 
membrane systems and reaction systems on the other hand.
In particular, we are interested in the benefits that can result from
establishing strong semantical links between the latter two models and
Petri nets. 
Our aim is to enhance the Petri net model in order to 
faithfully model the dynamics of the biological phenomena/processes 
represented by membrane systems and reaction systems.

After introducing Petri nets, we will   
outline how to understand and formalise their causality and concurrency
semantics. 

Then we turn to membrane systems.
Like membrane systems, Petri nets are in
essence
multiset rewriting systems. Using this key commonality we describe a faithful
translation from basic membrane systems to Petri nets. To capture the
compartmentalisation of membrane systems, the Petri net 
model has to be extended with localities which in turn leads to the idea of
locally synchronised executions.
In the thus extended model the standard causality semantics is no longer sound, 
and we will discuss possible solution to this problem.

Next we describe reaction systems which are 
a recently proposed model aimed at
investigating processes carried by biochemical reactions. Now, the resulting
computational model is remarkably different since in
reaction systems, biochemical reactions are modeled using a qualitative
rather than a quantitative approach.
As a consequence, counting  --- and hence the
multiset based calculus implemented in Petri nets ---
is no longer appropriate.
This insight leads to a new class of Petri nets, called set-nets, a novel and challenging
class of nets with intriguing (and yet to be discovered) properties.

We conclude the talk by demonstrating how in turn set-nets with localities
correspond to membrane systems with qualitative evolution 
rules.  

Altogether this talk aims to demonstrate the fruitful two-way
interaction between biological models and Petri nets.
Both membrane systems and reaction systems have inspired the introduction of 
new and relevant extensions to the basic net model, whereas having a Petri net
semantics opens the way for a new understanding, analysis and synthesis
techniques for biologically inspired systems.

The presentation is essentially self-contained; in  particular, all the
necessary details concerning   
the three models will be provided.

\bibliographystyle{mecbic} 

\end{document}